# Evaluation of reference equations of state for density prediction in regasified LNG mixtures using high-precision experimental data

Daniel Lozano-Martín[1], Dirk Tuma[2], and César R. Chamorro[3,*]

[1] GETEF, Universidad de Valladolid, Spain

[2] BAM Bundesanstalt für Materialforschung und -prüfung, Berlin, Germany

[3] MYER, Universidad de Valladolid, Spain

[*]cesar.chamorro@uva.es

## Abstract

This study evaluates the performance of three reference equations of state (EoS), AGA8-DC92, GERG-2008, and SGERG-88, in predicting the density of regasified liquefied natural gas (RLNG) mixtures. A synthetic nine-component RLNG mixture was gravimetrically prepared. High-precision density measurements were obtained using a single-sinker magnetic suspension densimeter over a temperature range of (250 to 350) K and pressures up to 20 MPa. The experimental data were compared with EoS predictions to evaluate their accuracy.

AGA8-DC92 and GERG-2008 showed excellent agreement with the experimental data, with deviations within their stated uncertainty. In contrast, SGERG-88 exhibited significantly larger deviations for this RLNG mixture, particularly at low temperatures of (250 to 260) K, where discrepancies reached up to 3 %. Even at 300 K, deviations larger than 0.4 % were observed at high pressures, within the model's uncertainty, but notably higher than those of the other two EoSs.

The analysis was extended to three conventional 11-component natural gas mixtures (labeled G420 NG, G431 NG, and G432 NG), previously studied by our group using the same methodology. While SGERG-88 showed reduced accuracy for the RLNG mixture, it performed reasonably well for these three mixtures, despite two of them have a very similar composition to the RLNG. This discrepancy is attributed to the lower $CO_2$ and $N_2$ content typical in RLNG mixtures, demostrating the sensitivity of EoS performance to minor differences in composition. These findings highlight the importance of selecting appropriate EoS models for accurate density prediction in RLNG applications.

## Highlights

Liquefied Natural Gas, LNG, adds flexibility to international natural gas trade.

A reference-quality Regasified LNG, RLNG, mixture was prepared by gravimetry.

Densities of the RLNG mixture were measured over a wide $T$ and $p$ range.

The experimental data were compared with the GERG-2008, AGA8-DC92, and SGERG-88 EoS.

SGERG-88 deviates notably from measured Regasified LNG mixture densities.

## Author contributions: CRediT

D.L-M., Data curation, Formal analysis, Investigation, Writing – review and editing. D.T., Data curation, Formal analysis, Investigation, Resources, Writing – review and editing. C.R.C., Investigation, Methodology, Resources, Supervision, Writing – original draft, Writing – review and editing.

## Acknowledgments

This study was supported by the Regional Government of Castilla y León (Junta de Castilla y León), the Ministry of Science and Innovation MCIN, and the European Union NextGenerationEU/PRTR, under project C17.I01.P01.S21, and by the European Metrology Programme for Innovation and Research (EMPIR), Funder ID: 10.13039/100014132, Grant No. 19ENG03 MefHySto.

This work represents the final experimental contribution utilizing the SSMSD at the University of Valladolid under the supervision of C.R. Chamorro. He extends his heartfelt gratitude to all coauthors of the 20 publications featuring high-precision density data for binary and multicomponent mixtures over the past two decades. Special thanks are due to Professor M.A. Villamañán for providing the opportunity to embark on this research journey, to the late Professor W. Wagner (†2024) and his team at the Ruhr-University of Bochum RUB, Germany, who generously hosted him in 1996 and introduced him to the fundamentals of this technique, and to his doctoral students M.E. Mondéjar (†2021), R. Hernández-Gómez, and D. Lozano-Martín for their dedication and invaluable collaboration.

## Statements and Declarations

The authors have no competing interests to declare that are relevant to the content of this article.

## 1. Introduction

Natural gas plays a vital role in current energy policies and is projected to become even more significant in the near future. Several factors contribute to this anticipated growth, including its relatively low cost, reduced carbon dioxide emissions compared to other fossil fuels, and its compatibility with renewable alternatives, such as hydrogen and biomethane [1,2]. These characteristics position natural gas as a key enabler in the ongoing transition toward a low-carbon energy system [3,4].

Liquefied Natural Gas (LNG) enhances the flexibility of global natural gas markets by allowing for a broader diversification of supply sources and offering alternatives against potential supply disruptions [5]. Natural gas (NG) primarily consists of methane, along with varying amounts of other hydrocarbons—such as ethane, propane, butane, and pentane—and non-combustible compounds, often referred to as impurities, like carbon dioxide, water, and nitrogen [6]. NG composition depends on the source and process of extraction. The liquefaction of natural gas improves its transportability, making economically feasible to move large volumes over long distances [7]. However, during LNG transport, some heat entry into the storage tanks is inevitable. This causes a small portion of the LNG to evaporate, enriching the vapor phase with the more volatile components and changing the composition of the remaining liquid. The evaporated portion, known as boil-off gas, is often used as fuel by the LNG carrier itself [8]. As a result, Regasified Liquefied Natural Gas (RLNG) typically exhibits a slightly different composition than the original NG, with a higher concentration of heavier hydrocarbons and fewer impurities [9,10].

For custody transfer and billing purposes, the accurate measurement of NG energy supplied is essential. Two main methodologies are employed to determine this value [11]. The first, known as the Gas Chromatography (GC) method, utilizes temperature, pressure, actual volumetric flow rate, and gas composition—determined via a process gas chromatographic analysis—as input parameters. Based on the gas composition, the higher heating value (*HHV*) can be calculated using the procedure defined in ISO 6976 [12]. The gas flow rate measured under flowing conditions must then be converted to standard conditions using appropriate equations of state (EoS), such as AGA8-DC92 [13] (referenced in ISO 12213-2 [14]) and GERG-2008 [15] (ISO 20765-2 [16]), which are widely adopted for this purpose.

The alternative approach, referred to as the Calorimetry Method (CM), also requires temperature, pressure, and actual volumetric flow rate as inputs. However, instead of relying on gas composition analysis, it employs the *HHV* obtained from an online calorimeter. The conversion of gas volume from flowing to standard conditions is performed using the SGERG-88 equation of state [17] (as specified in ISO 12213-3 [18]), which requires three of the following four parameters as inputs: *HHV*, relative density (also denoted as specific gravity, *SG*), and the individual concentrations of $CO_2$ and $N_2$. The CM does not require a detailed composition analysis, as it treats the hydrocarbon content as an equivalent hydrocarbon mixture.

While the GC method is well-established and considered highly accurate, it also has some limitations. One among these is the relatively slow response time of typical gas chromatographs, which, depending on the configuration of the setup, requires up to 10 minutes to produce a complete composition analysis, while flow meters provide flow measurements every second. The CM method offers a potential solution to this limitation and may serve as a viable alternative in natural gas energy metering systems, particularly with the advent of advanced calorimeters and their expanding role in the industry.

Currently, the AGA8-DC92 [13], GERG-2008 [15], and SGERG-88 [17] EoS are widely used by pipeline operators in both the United States and Europe for custody transfer and pipeline metering, as reported in various studies [19] and evidenced by their adoption in international standards [14,16,18]. These models have been developed using consolidated experimental data from pure substances and binary mixtures. However, their accuracy in predicting the behavior of ternary and more complex gas mixtures remains to be thoroughly validated. Therefore, further evaluation using high-quality experimental data is necessary which in turn requires the availability of highly accurate gas mixtures [3]. The primary objective of the present study is to assess and compare the performance of these three EoS in predicting the density of a RLNG mixture, which has been gravimetrically prepared to ensure minimal uncertainty in its composition.

To evaluate the performance of various reference equations of state (EoS) in predicting the density of regasified liquefied natural gas (RLNG) mixtures, a representative 9-component high-calorific natural gas mixture was prepared using gravimetric methods. The density of this mixture was measured with a single-sinker magnetic suspension densimeter (SSMSD) across a temperature range of (250 to 350) K and at pressures up to 20 MPa. The resulting experimental data were compared against three widely adopted reference EoS models in the natural gas industry: AGA8-DC92 [13], GERG-2008 [15], and SGERG-88 [17], all of which are commonly used for custody transfer and billing applications.

## 2. Theory and calculation

The AGA8-DC92 model [13] was developed by the American Gas Association and is recognized as a standard for natural gas property calculations. Initially based on a virial expansion of the compressibility factor, it was later reformulated into an explicit Helmholtz energy framework to better capture both calorific and volumetric properties [20] following multiple revisions. Its currently validated range includes gas and supercritical phases between (250 and 350) K and pressures up to 30 MPa, with an estimated uncertainty of 0.1 %.

The GERG-2008 EoS, developed by Kunz and Wagner [15] for the Groupe Européen de Recherches Gazières as an expansion of the GERG-2004 [21] EoS, extends the capabilities of AGA8-DC92 [13] by incorporating vapor-liquid equilibrium and liquid-phase behavior over a broader range of (60 to 700) K and pressures up to 70 MPa [16]. The GERG-2008 can model thermophysical properties of NG mixtures containing up to 21 components under pipeline conditions. Recent improvements of GERG-2008 have expanded its applicability in two important areas: (1) LNG applications, through improved departure functions for binary mixtures containing methane, increasing its accuracy in the subcooled liquid region (90 K to 180 K, up to 10 MPa) [22–24]; and (2) hydrogen-rich mixtures, by incorporating updated pure-component equations and improved binary interaction terms [25], supporting its use for hydrogen-enriched NG mixtures. Its estimated uncertainty in density for the temperature, pressure, and composition ranges considered in this work is 0.1 %.

Both AGA8-DC92 [13] and GERG-2008 [15] require a full compositional analysis of the gas mixture to reliably compute thermodynamic properties. In contrast, the SGERG-88 model [17], introduced in the 1997 release of ISO 12213-3 [18], uses a simplified input set consisting of any three parameters out of the higher heating value (*HHV*), relative density, and concentrations of $CO_2$ and $N_2$, plus mole fraction of $H_2$, at presence of $H_2$. SGERG-88 equation of state [26] is a virial-type thermal equation based on the Master (or Molar) GERG-88 virial equation [27]. Like all virial equations, it is only applicable in homogeneous gas phase, being adopted as a standard for the calculation of the compressibility factor, $Z$, of natural gas-like mixtures with an estimated uncertainty below 0.2 % within the pressure and temperature ranges of (0 to 12) MPa and (263 to 338) K, respectively. The uncertainty increases to about 0.5 % for pressures between (12 and 16) MPa or for temperatures outside the aforementioned range and exceeds 0.5 % for pressures above 16 MPa [18]. Internally, the SGERG-88 EoS treats the natural gas mixture as a five-component mixture consisting of nitrogen, carbon dioxide, hydrogen, carbon monoxide, and an equivalent hydrocarbon (representing all the alkane hydrocarbons of the mixture as a single pseudo-component with the same thermodynamic properties). This model offers a more accessible approach for estimating gas properties when detailed compositional data are unavailable.

## 2.1. RLNG mixture preparation

A synthetic natural gas mixture composed of nine components, representative of a typical RLNG composition with high calorific value, was prepared at the Federal Institute for Materials Research and Testing (BAM, Berlin, Germany) in a 10 $dm^3$ aluminum cylinder (Luxfer Gas Cylinders Inc., BAM cylinder no. 96054928-161019). The preparation followed the gravimetric method outlined in ISO 6142-1 [28] for reference materials, which ensures minimal uncertainty in the final composition. The mixture was prepared from high-purity individual gases through a series of intermediate pre-mixtures. Mass measurements along the entire filling procedure were performed using an electronic comparator balance (Sartorius LA 34000P-0CE) and a high-precision mechanical balance (Voland HCE 25), ensuring traceability and accuracy. After preparation, the gas mixture was homogenized by controlled rolling and heating. The final molar composition, $x_i$, together with the associated expanded ($k$ = 2) uncertainty in absolute terms, $U(x_i)$, is presented in Table 1.

**Table 1.** Composition of the RLNG mixture (cylinder no. 96054928-161019) studied in this work, with impurities compounds marked in italic type, and normalized composition without impurities.

| Component | Composition of the RLNG mixture | | Normalized composition of the RLNG mixture without impurities | |
|---|---|---|---|---|
| | $10^2\, x_i$ / mol·mol$^{-1}$ | $10^2\, U(x_i)$ / mol·mol$^{-1}$ | $10^2\, x_i$ / mol·mol$^{-1}$ | $10^2\, U(x_i)$ / mol·mol$^{-1}$ |
| Methane | 87.5790 | 0.0036 | 87.5791 | 0.0036 |
| Nitrogen | 0.11947 | 0.00015 | 0.11947 | 0.00015 |
| Carbon Dioxide | 0.020187 | 0.000082 | 0.020187 | 0.000082 |
| Ethane | 9.9437 | 0.0011 | 9.9437 | 0.0011 |
| Propane | 1.99856 | 0.00077 | 1.99857 | 0.00077 |
| *n*-Butane | 0.150132 | 0.000078 | 0.150132 | 0.000078 |
| Isobutane | 0.148984 | 0.000036 | 0.148984 | 0.000036 |
| *n*-Pentane | 0.019900 | 0.000024 | 0.019900 | 0.000024 |
| Isopentane | 0.020023 | 0.000024 | 0.020023 | 0.000024 |
| *Oxygen* | 0.000011 | 0.000009 | | |
| *Hydrogen* | 0.000010 | 0.000005 | | |
| *Carbon Monoxide* | 0.000001 | 0.000001 | | |
| *Neopentane* | 0.000042 | 0.000021 | | |
| *n-Hexane* | 0.0000005 | 0.0000004 | | |
| *Propene* | 0.000002 | 0.000002 | | |
| *Ethylene* | 0.0000015 | 0.0000013 | | |
| *Nitric Oxide* | 0.000000001 | 0.000000001 | | |

Following homogenization, the mixture was shipped to the University of Valladolid (Valladolid, Spain). Prior to shipment, the composition was independently verified at BAM using gas chromatography (GC) on a Siemens MAXUM II multichannel process analyzer. The analysis employed a bracketing calibration method in accordance with ISO 12963 [29], using certified reference gases of appropriate composition and following the BAM certification protocol. Further methodological details are available in a previous work [30] and references cited therein.

The uncertainty in the mole fractions of each component was assessed using the law of propagation of uncertainty, as recommended by the Guide to the Expression of Uncertainty in Measurement [31]. This evaluation considered the purity of the source gases and the entire preparation process. The agreement between the gravimetric and chromatographic compositions was within the acceptance limits defined by BAM, thereby the validation equals a successful certification of the mixture.

## 2.2. Experimental setup and procedure

Density measurements were performed at the University of Valladolid using a single-sinker magnetic suspension densimeter (SSMSD), known for its high accuracy across wide temperature and pressure ranges. The system consists of a measuring cell filled with the sample gas, where a monocrystalline silicon sinker with a precisely calibrated volume ($V_S$ = 226.4440 ± 0.0026 cm³ at ambient conditions) is suspended.

Density is determined based on Archimedes' principle, using the sinker's mass difference in vacuum and in the fluid, measured with a high-precision microbalance (XPE205DR, Mettler Toledo) via magnetic coupling:

$$\rho = \frac{(m_{S0} - m_{Sf})}{V_S(T,p)} \quad (1)$$

where $m_{S0}$ is the "true" mass of the sinker weighed in the evacuated measuring cell, $m_{Sf}$ is the "apparent" mass of the sinker weighed when the cell was filled with the fluid under study, and $V_S(T,p)$ is the volume of the sinker at temperature $T$ and pressure $p$.

The method, originally developed by Wagner's group at the University of Bochum, Germany [32–35], was adapted from two-sinker to single-sinker configurations [36,37], maintaining high precision especially at elevated densities. The measurement procedure involves the use of two calibrated masses of titanium and tantalum of nearly the same volume and whose difference in mass, due to the big difference in density, approximates the mass of the sinker. The alternating use of these masses allows the balance to operate near zero and minimizes linearity errors.

Corrections for force transmission errors—both apparatus- and fluid-specific—are applied. The apparatus-specific effect is determined by calculating the sinker weight in vacuum once all the data for an isotherm has been collected. This correction must always be applied to avoid significant errors [38]. The fluid-specific effect depends on the specific magnetic susceptibility of the fluid $\chi_s$, and on the so-called apparatus-specific constant $\varepsilon_\rho$, previously determined for our densimeter [39].

Pressure is monitored using two quartz transducers (Digiquartz 2300A-101 and Digiquartz 43KR-HHT-101, both from Paroscientific Inc), with expanded ($k$ = 2) uncertainties of $U(p) = [6.0 \cdot 10^{-5}(p/\text{MPa}) + 2 \cdot 10^{-3}]$ MPa for the low-pressure transducer (0 to 3) MPa, and $U(p) = [7.5 \cdot 10^{-5}(p/\text{MPa}) + 4 \cdot 10^{-3}]$ MPa for the high-pressure transducer (3 to 20) MPa.

Temperature control is achieved via an oil thermal bath (Dyneo DD-1000F, Julabo GmbH) and an electrical heating cylinder with a temperature controller (MC-E, Julabo GmbH), with measurements taken using platinum resistance thermometers (SPRT-25, Minco Products Inc.) and an AC resistance bridge (ASL F700, Automatic Systems Laboratory), yielding an expanded ($k$ = 2) uncertainty $U(T)$ = 0.015 K.

Further details on the experimental setup and methodology are displayed in prior publications [40,41].

### 2.3. Experimental uncertainty budget

The overall expanded ($k = 2$) uncertainty $U_T(\rho_{exp})$ for the experimental density measurements is summarized in Table 2, both in absolute and relative terms. This uncertainty incorporates contributions from the density determination itself, $U(\rho_{exp})$, previously characterized for our SSMSD as a function of fluid density, $\rho_{exp}$, and magnetic susceptibility, $\chi_s$ [39,41]:

$$U(\rho_{exp})/(\mathrm{kg \cdot m^{-3}}) = 2.5 \cdot 10^{4} \cdot \chi_s/(\mathrm{m^3 \cdot kg^{-1}}) + 1.1 \cdot 10^{-4} \cdot \rho_{exp}/(\mathrm{kg \cdot m^{-3}}) + 2.3 \cdot 10^{-2} \quad (1)$$

Additional sources of uncertainty — pressure, $u(p)$, temperature, $u(T)$, and composition, $u(x_i)$ — were combined using the law of propagation of uncertainty [31]:

$$U_T(\rho_{exp}) = 2\left[u(\rho_{exp})^2 + \left(\frac{\partial \rho}{\partial p}\Big|_{T,x} u(p)\right)^2 + \left(\frac{\partial \rho}{\partial T}\Big|_{p,x} u(T)\right)^2 + \sum_i \left(\frac{\partial \rho}{\partial x_i}\Big|_{T,p,x_j \neq x_i} u(x_i)\right)^2\right]^{0.5} \quad (3)$$

Partial derivatives of density with respect to pressure and temperature were estimated using REFPROP 10 [42], which employs the enhanced GERG-2008 EoS [22,25]. A detailed description of the REFPROP software can be found in [43]. Among all contributors to the uncertainty budget, uncertainties from pressure and density measurements dominate, reaching up to 0.097 kg·m⁻³ (0.37 %). In contrast, uncertainties from composition and temperature are significantly lower, below 0.018 kg·m⁻³ (0.01 %) and 0.0027 kg·m⁻³ (0.002 %), respectively. The total expanded uncertainty ranges from (0.033 to 0.11) kg·m⁻³, corresponding to relative uncertainties between 0.027 % and 0.52 %.

**Table 2.** Contributions to the expanded ($k = 2$) overall uncertainty in density, $U_T(\rho_{exp})$, for the RLNG mixture studied in this work.

| Source | Contribution ($k = 2$) | Units | Estimation in density ($k = 2$) | |
|---|---|---|---|---|
| | | | kg·m⁻³ | % |
| Temperature, $T$ | 0.015 | K | < 0.0027 | < 0.0019 |
| Pressure, $p$ | < 0.005 | MPa | 0.023 to 0.097) | 0.016 to 0.37 |
| Composition, $x_i$ | < 0.0004 | mol·mol⁻¹ | < 0.018 | < 0.010 |
| Density, $\rho$ | 0.024 to 0.051 | kg·m⁻³ | 0.024 to 0.051 | 0.021 to 0.37 |
| Sum | | | 0.033 to 0.11 | 0.027 to 0.52 |

## 3. Results

Density measurements were recorded at four temperatures, (250, 260, 300, and 350) K, with pressure successively decreasing in 1 MPa steps from 20 MPa down to 1 MPa. Figure 1 presents the data points for the mixture, along with the saturation curve calculated from the GERG-2008 EoS [15], the typical operational ranges for pipeline conditions in the gas industry, and the approved application limits for both the AGA8-DC92 [13] and the GERG-2008 EoS models.

**Figure 1.** ($p$, $T$)-phase diagram showing the experimental points measured (○) and the calculated phase envelope (solid line) using the GERG-2008 [15] EoS for the RLNG mixture. The marked temperature and pressure ranges represent the range of validity of the AGA8-DC92 [13] EoS (blue dotted line) and GERG-2008 EoS (red dashed line), and the area of interest for the gas industry (black dashed line).

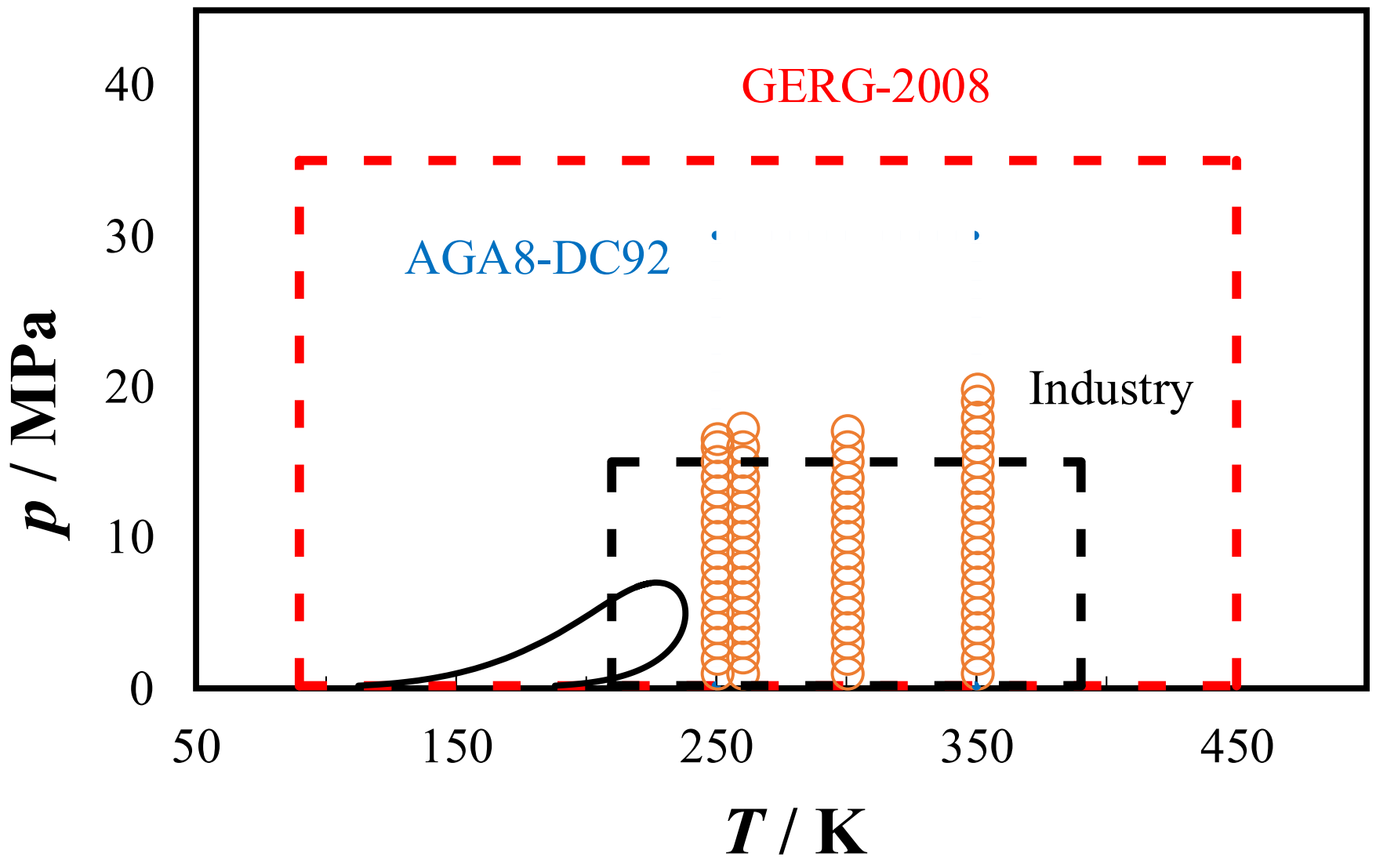

Table 3 presents the experimental ($p$, $\rho$, $T$) data for the RLNG mixture, along with the expanded uncertainty in density ($k$ = 2) calculated using Equation (2), expressed in both absolute and percentage terms. It also includes the relative deviations of the experimental densities from those predicted by the AGA8-DC92 [13], GERG-2008 [15], and SGERG-88 [17] EoS. The densities predicted by AGA8-DC92 and GERG-2008 were obtained using the REFPROP 10 software [42], while those predicted by the SGERG-88 EoS were calculated using the GasCalc software [44]. It is worth noting that, in the case of the GasCalc software for the SGERG-88 EoS, the input variables used in this work are composition, temperature, and pressure. The software internally converts this set of variables into the natural variables required by the SGERG-88 EoS.

**Table 3.** Experimental ($p$, $\rho_{exp}$, $T$) measurements for the RLNG mixture, absolute and relative expanded ($k$ = 2) uncertainty in density, $U(\rho_{exp})$, relative deviations from the density given by the AGA8-DC92 [13] EoS, $\rho_{AGA8\text{-}DC92}$, the GERG-2008 [15] EoS, $\rho_{GERG\text{-}2008}$, , and the SGERG-88 [17], $\rho_{SGERG\text{-}88}$

| $T$ / K[(a)] | $p$ / MPa[(a)] | $\rho_{exp}$ / (kg·m$^{-3}$[(a)]) | $U(\rho_{exp})$ / (kg·m$^{-3}$) | $10^2$ $U(\rho_{exp})$/ $\rho_{exp}$ | $10^2$ ($\rho_{exp}$ − $\rho_{AGA8\text{-}DC92}$)/$\rho_{AGA8\text{-}DC92}$ | $10^2$ ($\rho_{exp}$ − $\rho_{GERG\text{-}2008}$)/ $\rho_{GERG\text{-}2008}$ | $10^2$ ($\rho_{exp}$ − $\rho_{SGERG\text{-}88}$)/ $\rho_{SGERG\text{-}88}$ |
|---|---|---|---|---|---|---|---|
| | | | | 250 K | | | |
| 250.184 | 16.593 | 250.095 | 0.051 | 0.021 | 0.050 | 0.011 | – |
| 250.183 | 16.044 | 245.491 | 0.051 | 0.021 | 0.057 | 0.009 | – |
| 250.183 | 15.047 | 236.195 | 0.050 | 0.021 | 0.072 | 0.008 | 1.8 |
| 250.181 | 14.033 | 225.238 | 0.049 | 0.022 | 0.088 | 0.013 | 2.3 |
| 250.182 | 13.038 | 212.578 | 0.047 | 0.022 | 0.109 | 0.029 | 2.7 |
| 250.182 | 12.040 | 197.504 | 0.045 | 0.023 | 0.140 | 0.051 | 3.0 |
| 250.182 | 11.027 | 179.357 | 0.043 | 0.024 | 0.174 | 0.076 | 2.9 |
| 250.184 | 10.026 | 158.582 | 0.041 | 0.026 | 0.174 | 0.114 | 2.6 |
| 250.184 | 9.021 | 135.812 | 0.038 | 0.028 | 0.153 | 0.171 | 2.1 |
| 250.182 | 8.015 | 112.955 | 0.036 | 0.032 | 0.150 | 0.193 | 1.7 |
| 250.184 | 7.012 | 91.783 | 0.033 | 0.036 | 0.158 | 0.173 | 1.3 |
| 250.182 | 6.008 | 73.078 | 0.031 | 0.043 | 0.136 | 0.125 | 1.0 |
| 250.183 | 5.006 | 56.832 | 0.029 | 0.052 | 0.102 | 0.087 | 0.74 |
| 250.183 | 4.008 | 42.734 | 0.028 | 0.065 | 0.062 | 0.059 | 0.52 |
| 250.181 | 2.996 | 30.155 | 0.026 | 0.087 | 0.030 | 0.044 | 0.33 |
| 250.183 | 2.003 | 19.158 | 0.025 | 0.131 | 0.020 | 0.045 | 0.20 |
| 250.186 | 1.003 | 9.151 | 0.024 | 0.261 | 0.036 | 0.059 | 0.12 |
| | | | | 260 K | | | |
| 260.179 | 17.278 | 234.440 | 0.050 | 0.021 | 0.083 | 0.030 | 0.37 |
| 260.180 | 16.046 | 223.151 | 0.048 | 0.022 | 0.091 | 0.035 | 0.94 |
| 260.179 | 15.043 | 212.601 | 0.047 | 0.022 | 0.100 | 0.043 | 1.3 |
| 260.180 | 14.049 | 200.690 | 0.046 | 0.023 | 0.114 | 0.055 | 1.6 |
| 260.178 | 13.040 | 186.904 | 0.044 | 0.024 | 0.127 | 0.064 | 1.8 |
| 260.178 | 12.041 | 171.432 | 0.042 | 0.025 | 0.135 | 0.078 | 1.8 |
| 260.179 | 11.021 | 153.931 | 0.040 | 0.026 | 0.134 | 0.106 | 1.6 |
| 260.179 | 10.026 | 135.646 | 0.038 | 0.028 | 0.120 | 0.125 | 1.3 |
| 260.177 | 9.021 | 116.864 | 0.036 | 0.031 | 0.117 | 0.133 | 1.0 |
| 260.178 | 8.014 | 98.607 | 0.034 | 0.035 | 0.119 | 0.125 | 0.83 |
| 260.178 | 7.010 | 81.648 | 0.032 | 0.039 | 0.114 | 0.108 | 0.65 |
| 260.176 | 6.011 | 66.280 | 0.030 | 0.046 | 0.094 | 0.086 | 0.50 |
| 260.175 | 5.010 | 52.388 | 0.029 | 0.055 | 0.066 | 0.064 | 0.37 |
| 260.176 | 4.007 | 39.846 | 0.027 | 0.069 | 0.041 | 0.050 | 0.26 |
| 260.173 | 3.004 | 28.510 | 0.026 | 0.092 | 0.023 | 0.043 | 0.18 |
| 260.174 | 2.004 | 18.211 | 0.025 | 0.137 | 0.017 | 0.042 | 0.11 |
| 260.180 | 1.003 | 8.752 | 0.024 | 0.273 | 0.036 | 0.058 | 0.08 |
| | | | | 300 K | | | |
| 300.098 | 17.088 | 165.989 | 0.042 | 0.025 | 0.087 | 0.035 | 0.16 |
| 300.096 | 16.018 | 155.831 | 0.041 | 0.026 | 0.082 | 0.036 | 0.30 |
| 300.097 | 15.011 | 145.778 | 0.039 | 0.027 | 0.077 | 0.039 | 0.39 |
| 300.097 | 14.011 | 135.398 | 0.038 | 0.028 | 0.072 | 0.041 | 0.44 |
| 300.098 | 13.012 | 124.704 | 0.037 | 0.030 | 0.069 | 0.044 | 0.45 |
| 300.099 | 12.012 | 113.786 | 0.036 | 0.032 | 0.067 | 0.046 | 0.43 |
| 300.099 | 11.013 | 102.815 | 0.035 | 0.034 | 0.070 | 0.053 | 0.39 |
| 300.099 | 10.014 | 91.870 | 0.033 | 0.036 | 0.061 | 0.049 | 0.34 |
| 300.096 | 9.011 | 81.053 | 0.032 | 0.040 | 0.052 | 0.046 | 0.28 |
| 300.097 | 8.007 | 70.488 | 0.031 | 0.044 | 0.044 | 0.045 | 0.23 |
| 300.097 | 7.007 | 60.291 | 0.030 | 0.049 | 0.039 | 0.046 | 0.19 |
| 300.097 | 6.005 | 50.463 | 0.029 | 0.057 | 0.031 | 0.043 | 0.14 |
| 300.097 | 5.004 | 41.045 | 0.028 | 0.067 | 0.025 | 0.041 | 0.11 |
| 300.098 | 4.003 | 32.050 | 0.027 | 0.083 | 0.023 | 0.041 | 0.08 |
| 300.098 | 2.990 | 23.360 | 0.026 | 0.109 | 0.020 | 0.038 | 0.06 |
| 300.097 | 2.002 | 15.277 | 0.025 | 0.161 | 0.034 | 0.050 | 0.05 |
| 300.102 | 1.002 | 7.471 | 0.024 | 0.318 | 0.040 | 0.052 | 0.05 |
| | | | | 350 K | | | |
| 350.095 | 19.859 | 141.435 | 0.039 | 0.028 | 0.018 | < 0.001 | −0.85 |
| 350.094 | 19.013 | 135.742 | 0.038 | 0.028 | 0.009 | −0.003 | −0.73 |
| 350.094 | 18.017 | 128.875 | 0.038 | 0.029 | −0.002 | −0.008 | −0.61 |
| 350.095 | 17.012 | 121.787 | 0.037 | 0.030 | −0.010 | −0.009 | −0.49 |
| 350.096 | 16.014 | 114.598 | 0.036 | 0.031 | −0.016 | −0.010 | −0.40 |
| 350.098 | 15.009 | 107.237 | 0.035 | 0.033 | −0.019 | −0.009 | −0.32 |
| 350.100 | 14.006 | 99.803 | 0.034 | 0.034 | −0.021 | −0.007 | −0.25 |
| 350.099 | 13.010 | 92.344 | 0.033 | 0.036 | −0.023 | −0.007 | −0.20 |
| 350.098 | 12.009 | 84.803 | 0.033 | 0.038 | −0.025 | −0.007 | −0.16 |
| 350.099 | 11.008 | 77.258 | 0.032 | 0.041 | −0.019 | 0.001 | −0.12 |
| 350.098 | 10.005 | 69.713 | 0.031 | 0.044 | −0.022 | −0.002 | −0.10 |
| 350.098 | 9.006 | 62.238 | 0.030 | 0.048 | −0.021 | −0.001 | −0.08 |
| 350.098 | 8.007 | 54.833 | 0.029 | 0.053 | −0.020 | < 0.001 | −0.06 |
| 350.098 | 7.006 | 47.508 | 0.028 | 0.060 | −0.016 | 0.002 | −0.05 |
| 350.098 | 6.005 | 40.293 | 0.027 | 0.068 | −0.017 | < 0.001 | −0.05 |
| 350.098 | 5.004 | 33.204 | 0.027 | 0.080 | −0.014 | 0.002 | −0.04 |
| 350.099 | 4.003 | 26.259 | 0.026 | 0.099 | −0.012 | 0.002 | −0.03 |
| 350.098 | 3.004 | 19.467 | 0.025 | 0.129 | −0.004 | 0.009 | −0.02 |
| 350.099 | 2.003 | 12.822 | 0.024 | 0.190 | 0.010 | 0.021 | −0.004 |
| 350.102 | 1.003 | 6.342 | 0.024 | 0.372 | 0.046 | 0.054 | 0.04 |

(a) Expanded uncertainties ($k = 2$): $U(p > 3)/\text{MPa} = 75 \cdot 10^{-6} \cdot \frac{p}{\text{MPa}} + 3.5 \cdot 10^{-3}$; $U(p < 3)/\text{MPa} = 60 \cdot 10^{-6} \cdot \frac{p}{\text{MPa}} + 1.7 \cdot 10^{-3}$; $U(T) = 15$ mK; $\frac{U(\rho)}{\text{kg·m}^{-3}} = 2.5 \cdot 10^4 \frac{\chi_S}{m^3 kg^{-1}} + 1.1 \cdot 10^{-4} \cdot \frac{\rho}{\text{kg·m}^{-3}} + 2.3 \cdot 10^{-2}$.

## 4. Discussion

### 4.1. Deviation analysis of EoS for the RLNG density data

Figure 2 presents the percentage relative deviations of the experimental density data for the RLNG mixture from the values calculated using the AGA8-DC92 [13], GERG-2008 [15], and SGERG-88 [17] EoS. The density values predicted by the different EoS were calculated using the normalized, impurity-free composition provided in Table 1. The numerical values of these deviations are listed in the last three columns of Table 3.

**Figure 2.** Relative deviations in density of experimental ($p$, $\rho_{exp}$, $T$) data of RLNG mixture from density values calculated from (a) AGA8-DC92 [13] EoS, $\rho_{AGA8\text{-}DC92}$, (b) GERG-2008 [15] EoS, $\rho_{GERG\text{-}2008}$, and (c) SGERG-88 [17], $\rho_{SGERG\text{-}88}$, as function of pressure for different temperatures: + 250 K, □ 260 K, △ 300 K, ○ 350 K. Dashed lines indicate the expanded ($k$ = 2) uncertainty of the corresponding EoS. Error bars on the 260 K data set indicate the expanded ($k$ = 2) uncertainty of the experimental density. Note the different scale on the y-axis in plot (c).

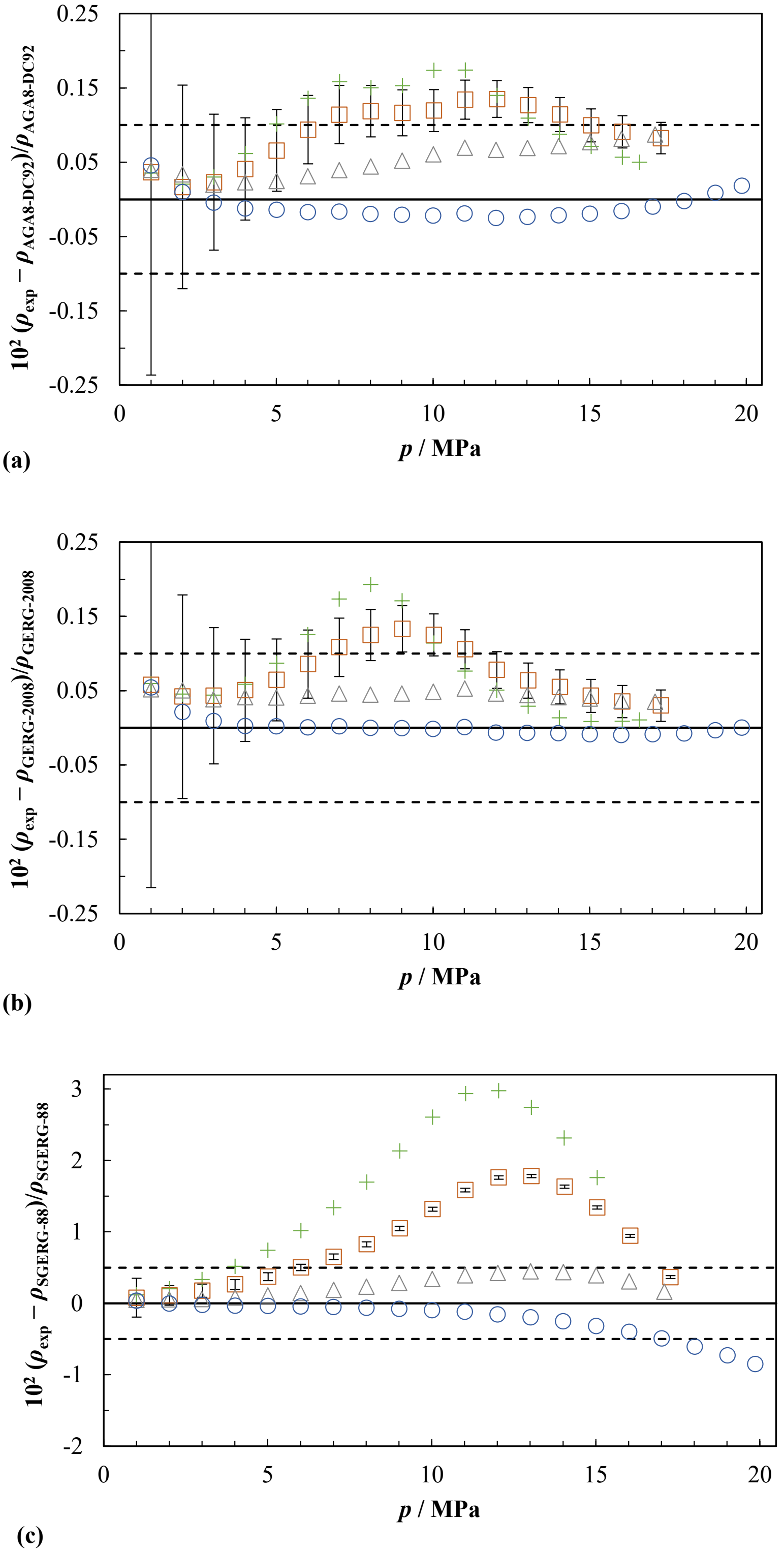

The relative deviations between the experimental density data and the AGA8-DC92 [13] and GERG-2008 [15] EoS, shown in Figures 2(a) and 2(b), respectively, generally fall within the stated uncertainty of these models ($U(\rho_{EoS})$ = 0.1 %), except at the lower temperatures of (250 and 260) K and pressures between (5 and 15) MPa, where deviations can reach up to +0.20 %.

In contrast, the relative deviations between the experimental data and the SGERG-88 [17] EoS, shown in Figure 2(c), are larger approximately by one order of magnitude, in the limits, or above, the stated uncertainty of this model. At 250 K and pressures between (11 and 12) MPa, deviations can reach up to as much as +3.0 %. At 260 K, deviations of up to +1.8 % are observed at pressures between (12 and 13) MPa. Only at 300 K do the deviations remain within an uncertainty of 0.5 % across all pressures investigated. However, at the highest measured temperature of 350 K, deviations again exceed this limit at pressures above 17 MPa, showing negative deviations of up to −0.85 %.

### 4.2. Deviation analysis of EoS for other NG density data

Three natural gas (NG) mixtures were previously studied by our group using the same experimental technique, and the results have already been published. The compositions of these three mixtures, designated as G420 NG [30], G431 NG [45], and G432 NG [46], along with the composition of the RLNG mixture investigated in this work, are presented in Table 4. All four mixtures were prepared gravimetrically at BAM, ensuring minimal uncertainty in their compositions.

**Table 4.** Normalized, impurity-free composition of the RLNG mixture studied in this work, and of the other natural gas mixtures from our previous studies using the same experimental technique: G420 NG [30], G431 NG [45], and G432 NG [46][(a)],together with the molar mass $M$, normalized density $\rho_n$, relative density $SG$, higher heating value $HHV$, and Wobbe index $W_s$, for the four gas mixtures estimated using the REFPROP 10 software.

| Component | RLNG | G420 NG | G431 NG | G432 NG |
|---|---|---|---|---|
| | $10^2\,x_i$ / mol·mol$^{-1}$ | $10^2\,x_i$ / mol·mol$^{-1}$ | $10^2\,x_i$ / mol·mol$^{-1}$ | $10^2\,x_i$ / mol·mol$^{-1}$ |
| Methane | 87.5790 | 87.6639 | 97.2362 | 85.0063 |
| Nitrogen | 0.11947 | 4.3215 | 1.40097 | 0.9508 |
| Carbon Dioxide | 0.020187 | 1.62267 | 0.36146 | 1.44823 |
| Ethane | 9.9437 | 4.2252 | 0.398705 | 8.99177 |
| Propane | 1.99856 | 1.04900 | 0.201221 | 3.00256 |
| *n*-Butane | 0.150132 | 0.212726 | 0.100398 | 0.19994 |
| Isobutane | 0.148984 | 0.210383 | 0.100431 | 0.200443 |
| *n*-Pentane | 0.019900 | 0.051811 | 0.100853 | 0.100089 |
| Isopentane | 0.020023 | 0.052238 | 0.049928 | 0.049929 |
| *n*-Hexane | – | 0.052611 | 0.049883 | 0.049965 |
| Oxygen | – | 0.537990 | – | – |

| Property | RLNG | G420 NG | G431 NG | G432 NG |
|---|---|---|---|---|
| Molar mass, $M$ / (g/mol) | 18.166 | 18.260 | 16.628 | 18.954 |
| Normalized density, $\rho_n$ / (kg·m$^{-3}$) | 0.77034 | 0.77400 | 0.70468 | 0.80379 |
| Relative density, $SG$ | 0.629 | 0.632 | 0.575 | 0.656 |
| Higher heating value, $HHV$ / (MJ·m$^{-3}$) | 42.003 | 37.678 | 37.749 | 41.726 |
| Wobbe index $W_s$ / (MJ·m$^{-3}$) | 52.979 | 47.411 | 49.783 | 51.523 |

[(a)] The uncertainties of the NG mixtures are given in the corresponding references.

Table 4 also includes the molar mass $M$, normalized density $\rho_n$, relative density $SG$, higher heating value $HHV$, and Wobbe index $W_s$, for the four gas mixtures. These properties were estimated using the REFPROP 10 software [42] based on the normalized, impurity-free compositions and under reference conditions of 288.15 K and 0.101325 MPa. Based on these values, all four mixtures can be classified as high-calorific natural gases (H-Gas). According to the European standard EN 437 [47], H-Gas mixtures are defined as those with a Wobbe index between 47.2 and 54.7 MJ·m$^{-3}$ and a relative density ($SG$, defined as the gas density relative to air) between 0.55 and 0.75.

The G420 NG mixture is an 11-component natural gas with a methane content comparable to that of the RLNG mixture, but with lower ethane and propane contents and higher concentrations of $N_2$ and $CO_2$ than the RLNG mixture. Densities were measured at five different temperatures (260, 275, 300, 325, and 350) K and up to a maximum pressure of 20 MPa. The G431 NG mixture is a 10-component natural gas primarily composed of methane (> 97 %), but, again, with higher $N_2$ and $CO_2$ contents than the RLNG mixture. Densities were measured at five different temperatures (250, 275, 300, 325, and 350) K and up to a maximum pressure of 20 MPa. The G432 NG mixture is also a 10-component natural gas, with methane, ethane, and propane contents similar to those of the RLNG mixture; however, it also contains significantly higher amounts of $N_2$ and $CO_2$ than the RLNG mixture. Densities were measured at five different temperatures (260, 275, 300, 325, and 350) K and up to a maximum pressure of 20 MPa.

Figures 3, 4, and 5 show the percentage relative deviations of the experimental density data from the values calculated using the AGA8-DC92 [13], GERG-2008 [15], and SGERG-88 [17] EoS for the G420 NG, G431 NG, and G432 NG mixtures, respectively. Comparisons with the AGA8-DC92 and GERG-2008 EoS were previously analyzed in earlier publications (G420 NG [30], G431 NG [45], and G432 NG [46]), and are included here for completeness. The comparison with the SGERG-88 EoS represents a new contribution of this work. Table 5 provides a statistical comparison of the experimental density data for the three natural gas mixtures, G420 NG, G431 NG, and G432 NG, relative to the three EoS models, along with the corresponding statistical values for the RLNG mixture.

**Table 5.** Statistical analysis of the ($p$, $\rho$, $T$) data set with respect to AGA8-DC92 [13], GERG-2008 [15], and SGERG-88 [17] EoS for the RLNG mixture studied in this work, and for other natural gas mixtures (G420 NG [30], G431 NG [45], and G432 NG [46]) from the literature. AARD = average absolute value of relative deviation, ASRD = average signed relative deviation, MaxARD = maximum absolute value of relative deviation.

| Reference | $N$[(a)] | Covered ranges | | Experimental vs AGA8-DC92EoS | | | Experimental vs GERG-2008 EoS | | | Experimental vs SGERG-88 EoS | | |
|---|---|---|---|---|---|---|---|---|---|---|---|---|
| | | $T$ / K | $p$ / MPa | AARD / % | ASRD / % | MaxARD / % | AARD / % | ASRD / % | MaxARD / % | AARD / % | ASRD / % | MaxARD / % |
| RLNG | 69 | 250 to 350 | 1 to 20 | 0.063 | 0.056 | 0.17 | 0.048 | 0.046 | 0.19 | 0.66 | 0.55 | 2.98 |
| G420 NG | 100 | 260 to 350 | 1 to 20 | 0.078 | −0.078 | 0.13 | 0.027 | -0.025 | 0.095 | 0.24 | −0.22 | 1.51 |
| G431 NG | 96 | 250 to 350 | 1 to 20 | 0.012 | −0.007 | 0.054 | 0.032 | 0.032 | 0.049 | 0.30 | −0.22 | 1.61 |
| G432 NG | 93 | 260 to 350 | 1 to 20 | 0.040 | −0.033 | 0.15 | 0.043 | −0.023 | 0.21 | 0.18 | −0.07 | 1.16 |

(a) Number of experimental points.

In Table 5, the statistical indicators are defined as follows: AARD (average absolute value of the relative deviations), ASRD (average signed relative deviations), and MaxARD (maximum absolute value of the relative deviations), as given by Eqs. (4) to (6):

$$\mathrm{AARD} = \frac{1}{N}\sum_{i=1}^{N}\left|10^2 \frac{\rho_{\mathrm{i,exp}}-\rho_{\mathrm{i,EoS}}}{\rho_{\mathrm{i,EoS}}}\right| \quad (4)$$

$$\mathrm{ASRD} = \frac{1}{N}\sum_{i=1}^{N}\left(10^2 \frac{\rho_{\mathrm{i,exp}}-\rho_{\mathrm{i,EoS}}}{\rho_{\mathrm{i,EoS}}}\right) \quad (5)$$

$$\mathrm{MaxARD} = \max\left|10^2 \frac{\rho_{\mathrm{i,exp}}-\rho_{\mathrm{i,EoS}}}{\rho_{\mathrm{i,EoS}}}\right| \quad (6)$$

The density values given by the different EoS were calculated using the normalized composition without impurities given in Table 4.

Figures 3(a) and 3(b) show the relative deviations between the experimental density data for the G420 NG mixture and the AGA8-DC92 [13] and GERG-2008 [15] EoS, respectively. Both models provide an excellent representation of the experimental data, with most values falling within their stated uncertainty limits, even at the lowest temperature considered in this study, i.e., 260 K (instead of the typical 250 K). Only three data points, at 260 K and pressures above 15 MPa, exhibit deviations exceeding −0.1 %. A comparison of the two models reveals that GERG-2008 performs slightly better (AARD of 0.027 %) than AGA8-DC92 (AARD of 0.078 %). The SGERG-88 [17] EoS shows lower accuracy, with an AARD of 0.23 % and a MaxARD of 1.51 %. Nevertheless, the largest part of the experimental data remains within the stated uncertainty of the SGERG-88 EoS, except at pressures above 15 MPa where experimental data show negative deviations, as illustrated in Figure 3(c).

**Figure 3.** Relative deviations in density of experimental ($p$, $\rho_{\mathrm{exp}}$, $T$) data of G420 NG [30] mixture from density values calculated from (a) AGA8-DC92 [13] EoS, $\rho_{\mathrm{AGA8\text{-}DC92}}$, (b) GERG-2008 [15] EoS, $\rho_{\mathrm{GERG\text{-}2008}}$, and (c) SGERG-88 [17], $\rho_{\mathrm{SGERG\text{-}88}}$, as function of pressure for different temperatures: □ 260 K, ◇ 275 K, △ 300 K, × 325 K, ○ 350 K. Dashed lines indicate the expanded ($k$ = 2) uncertainty of the corresponding EoS. Error bars on the 260 K data set indicate the expanded ($k$ = 2) uncertainty of the experimental density. Note the different scale on the y-axis in plot (c).

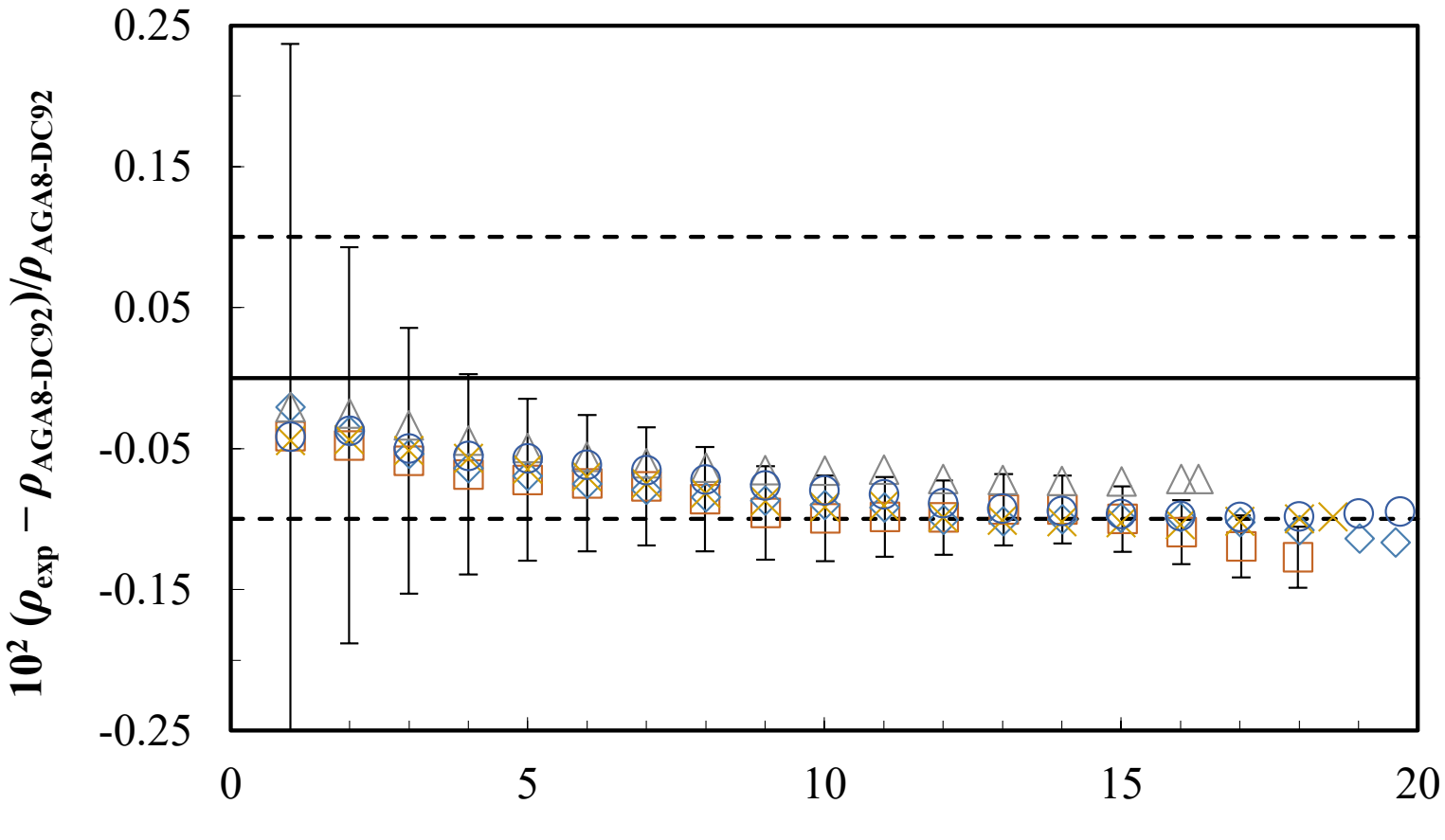

**(a)**

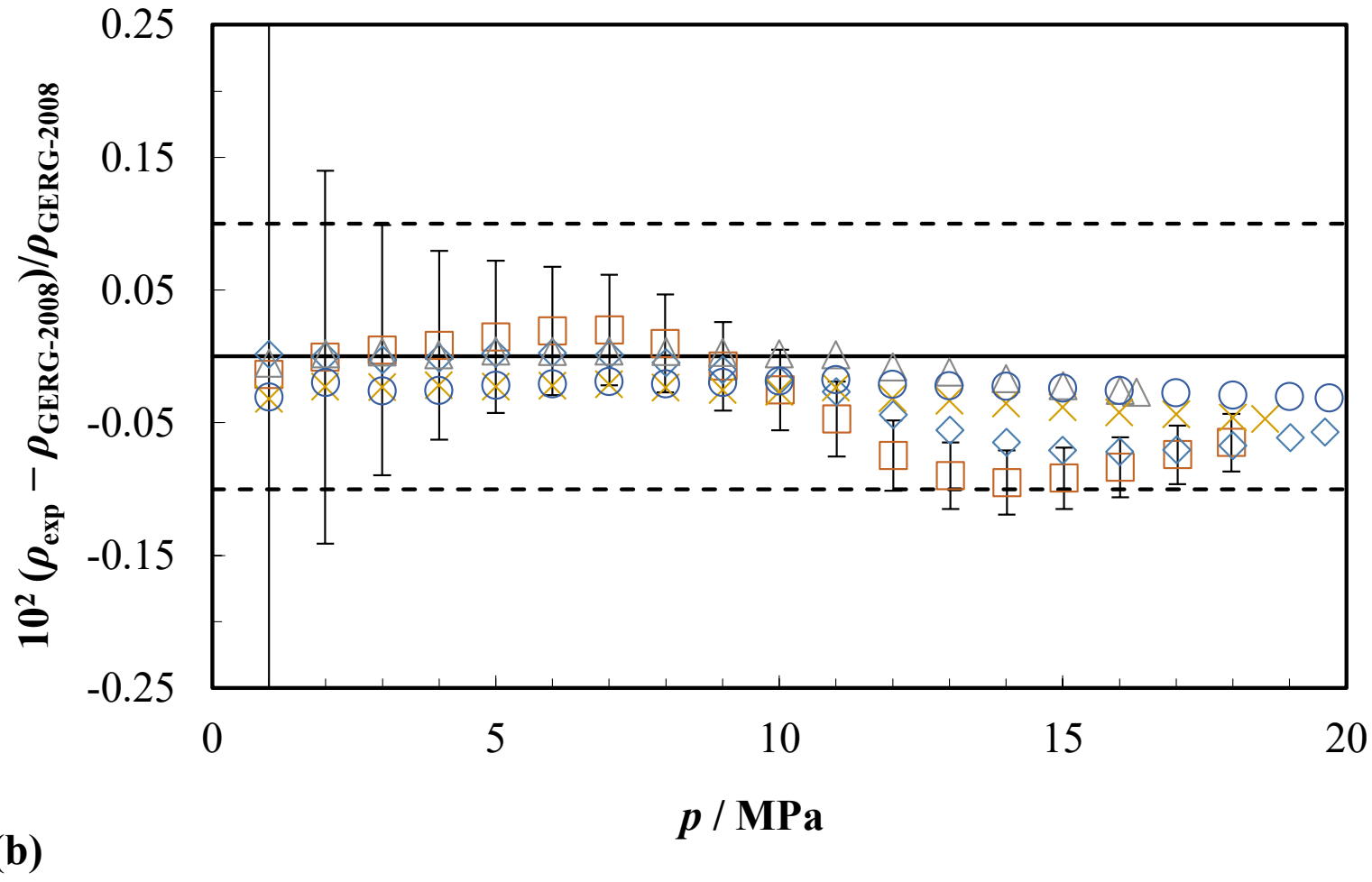

**(b)**

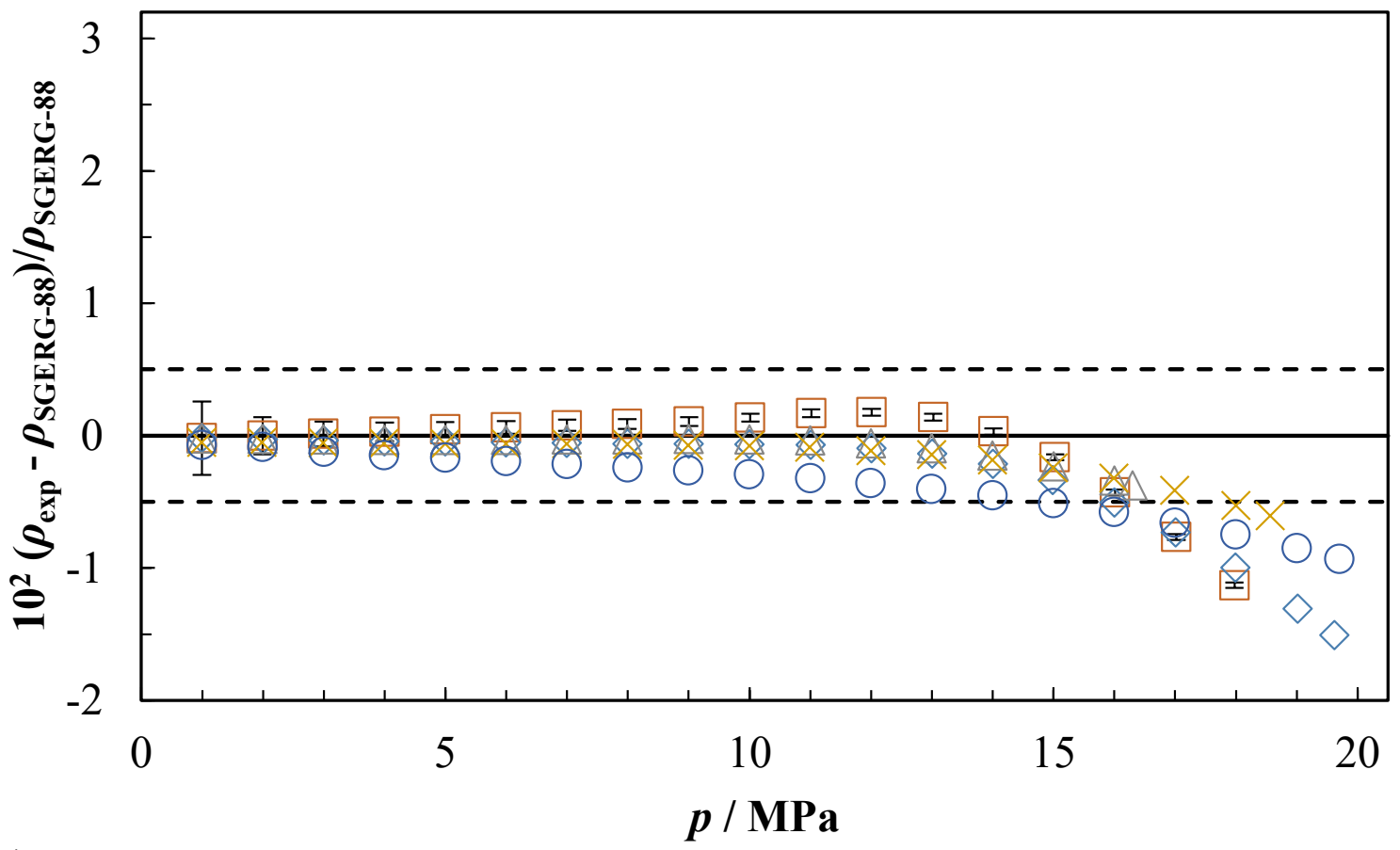

**(c)**

Figures 4 and 5 display the relative deviations between the experimental density data and the three EoS models used for comparison for the G431 NG and G432 NG mixtures, respectively. The behavior observed is very similar to that previously discussed for the G420 NG mixture in Figure 3. Both the AGA8-DC92 [13] and GERG-2008 [15] EoS perform very well in describing the experimental data, particularly for the G431 NG mixture, which has the highest methane content. For this mixture, the AARD values are 0.012 % for AGA8-DC92 and 0.032 % for GERG-2008. Only a few data points for the G432 NG mixture, at the lowest temperature of 250 K, show deviations exceeding the stated uncertainty of the EoS. The SGERG-88 [17] EoS again demonstrates slightly lower performance, with AARD values of 0.30 % for G431 NG and 0.18 % for G432 NG, and MaxARD values of 1.61 % and 1.16 %, respectively. Nonetheless, most of the experimental data are still captured within the uncertainty bounds of the SGERG-88 EoS, except at pressures above 15 MPa, where experimental data display again negative deviations, as shown in Figures 4(c) and 5(c).

**Figure 4.** Relative deviations in density of experimental ($p$, $\rho_{exp}$, $T$) data of G431 NG [45] mixture from density values calculated from (a) AGA8-DC92 [13] EoS, $\rho_{AGA8\text{-}DC92}$, (b) GERG-2008 [15] EoS, $\rho_{GERG\text{-}2008}$, and (c) SGERG-88 [17], $\rho_{SGERG\text{-}88}$, as function of pressure for different temperatures: + 250 K, ◇ 275 K, △ 300 K, × 325 K, ○ 350 K. Dashed lines indicate the expanded ($k$ = 2) uncertainty of the corresponding EoS. Error bars on the 250 K data set indicate the expanded ($k$ = 2) uncertainty of the experimental density. Note the different scale on the y-axis in plot (c).

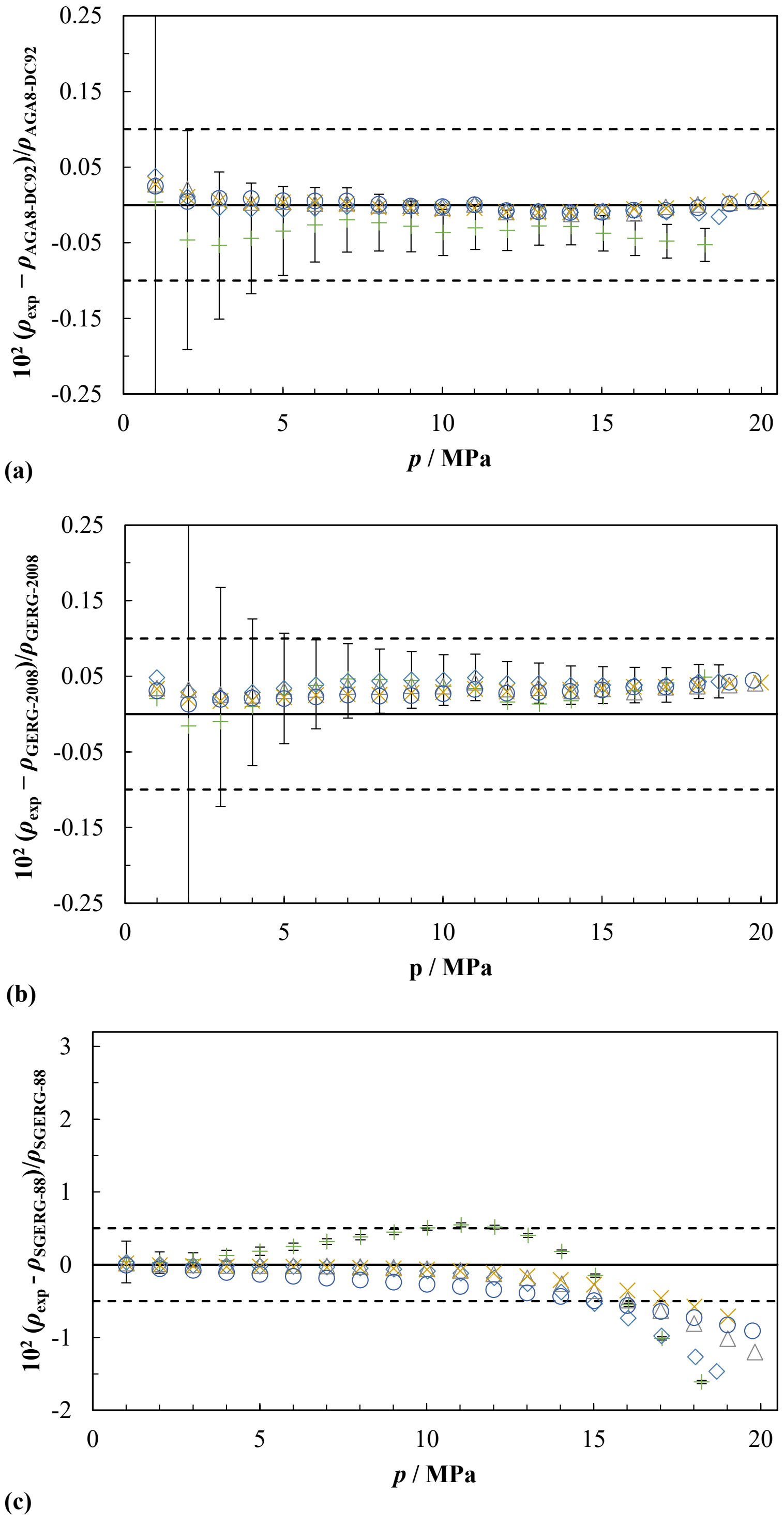

**Figure 5.** Relative deviations in density of experimental ($p$, $\rho_{exp}$, $T$) data of G432 [46] NG mixture from density values calculated from (a) AGA8-DC92 [13] EoS, $\rho_{AGA8\text{-}DC92}$, (b) GERG-2008 [15] EoS, $\rho_{GERG\text{-}2008}$, and (c) SGERG-88 [17], $\rho_{SGERG\text{-}88}$, as function of pressure for different temperatures: □ 260 K, ◇ 275 K, △ 300 K, × 325 K, ○ 350 K. Dashed lines indicate the expanded ($k$ = 2) uncertainty of the corresponding EoS. Error bars on the 260 K data set indicate the expanded ($k$ = 2) uncertainty of the experimental density. Note the different scale on the y-axis in plot (c).

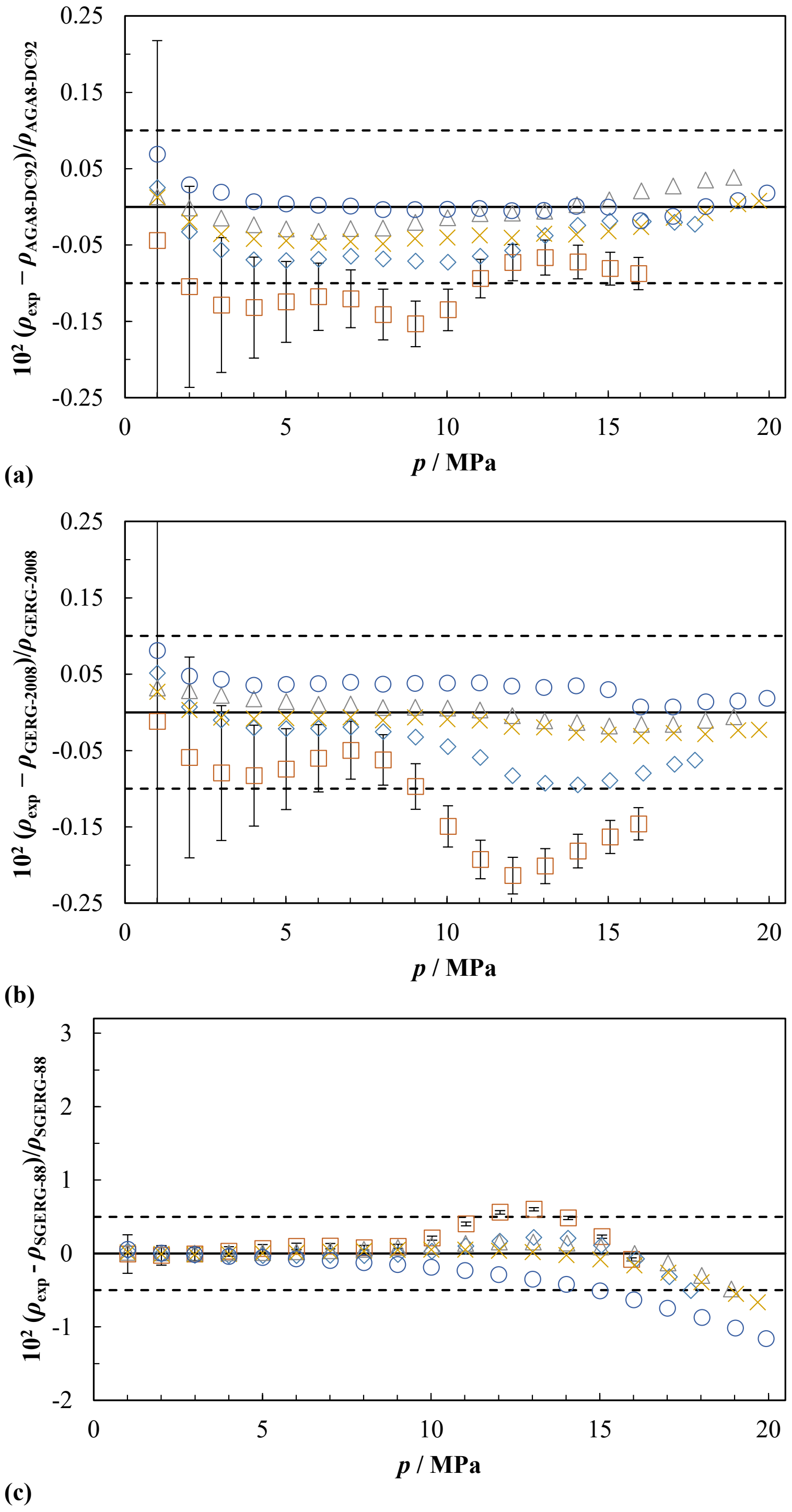

The performance of the AGA8-DC92 [13] and GERG-2008 [15] EoS has also been evaluated by other authors for various natural gas mixtures. For instance, Farzaneh-Gord et al. [48] compared the AGA8-DC92 and GERG-2008 EoS for five typical Iranian natural gas compositions. Their study demonstrated that GERG-2008 consistently outperformed AGA8-DC92 across the entire range of pressures and temperatures considered. Notably, none of the mixtures studied was of the RLNG type. The results also indicated that GERG-2008 tends to predict higher compressibility factors than AGA8-DC92 in the practical measurement range.

Ahmadi et al. [49] simultaneously measured the density and speed of sound of a synthetic natural gas mixture (~88 mol-% methane) These measurements were carried out along five isotherms at temperatures between (323 and 415) K and pressures up to the remarkably high value of 56 MPa. The experimental results showed good agreement with the predictions of both the AGA8-DC92 [13] and GERG-2008 [15] EoS.

The GERG-2008 [15] EoS was compared with the cubic equations of Peng–Robinson and Redlich–Kwong–Soave and the equation of state of Lee–Kesler–Plöcker for the representation of thermodynamic properties of natural gas in another study [9]. The EoS showed high potential for accurate process-modelling.

In another study, Chaczykowski [19] analyzed the performance of the SGERG-88 [17] and AGA8-DC92 [13] EoS for a nine-component natural gas mixture with a methane content greater than 98 mol-%. The results indicated that both models yielded similar predictions.

## 5. Conclusions

A nine-component synthetic natural gas mixture, representative of typical RLNG, was prepared in reference quality with minimal uncertainty in composition at the Federal Institute for Materials Research and Testing (BAM, Berlin, Germany). High-precision experimental density measurements for this mixture were obtained using a single-sinker magnetic suspension densimeter (SSMSD) at the University of Valladolid (Valladolid, Spain). The experimental densities were compared with the densities calculated from three equations of state (EoS) commonly used in the natural gas industry: AGA8-DC92 [13], GERG-2008 [15], and SGERG-88 [17]. This comparison was also analyzed for three other NG mixtures, G420 NG, G431 NG, and G432 NG, previously measured using the same experimental technique by our group.

Both the AGA8-DC92 [13] and GERG-2008 [15] EoS demonstrated an excellent agreement with experimental data for the RLNG mixture and for the G420 NG, G431 NG, and G432 NG mixtures. The SGERG-88 [17] EoS showed a slightly lower accuracy for the three NG mixtures, but still within its stated uncertainty, but it failed to accurately predict the density of the RLNG mixture at the lowest temperatures studied (250 and 260) K, where deviations of up to +3 % were observed. Even at near-ambient conditions (e.g., 300 K), deviations reached up to +0.4 % at pressures between (11 and 15) MPa, within the claimed uncertainty of the SGERG-88 EoS, but nearly one order of magnitude greater than those observed for the other two EoS.

The compositions of the G420 NG, G431 NG, and G432 NG differ considerably in methane content, ranging from 87.7 mol-% in G420 NG to 97.2 mol-% in G431 NG, and in the combined ethane and propane content, ranging from 0.6 mol-% in G431 NG to 12.0 mol-% in G432 NG. Nevertheless, the SGERG-88 [17] EoS shows a similar predictive capacity for all three mixtures. It is worth noting that, despite the apparent similarity in composition between RLNG and G432 NG (similar combined ethane and propane content, i.e., around 12 mol-%) and between RLNG and G420 NG (similar methane content, i.e., around 87.6 mol-%), the capacity of the SGERG-88 EoS to predict their densities differs significantly. While the densities of G432 NG and G420 NG were well represented by the SGERG-88, the RLNG densities at low temperatures showed considerable deviations from the predicted values. The main difference between these mixtures lies in their $CO_2$ and $N_2$ contents, as RLNG typically contains lower concentrations of both components. The $CO_2$ content in RLNG mixtures is generally below 0.2 mol-%, often in the range of (0.01 to 0.1) mol-%, and $N_2$ is typically below 1 mol-%, often around (0.1 to 0.5) mol-%. This characteristic is clearly reflected in the RLNG mixture, but not in the G420 NG, G431 NG, and G432 NG mixtures, which contain more than 0.3 mol-% $CO_2$ and more than 0.95 mol-% $N_2$. These composition differences may explain the discrepancies observed in density predictions by the SGERG-88 EoS.